\documentclass[aip,jcp %
 aip,
 amsmath,amssymb,
 reprint,%
]{revtex4-1}

\usepackage{graphicx}
\usepackage{dcolumn}
\usepackage{bm}

\usepackage[utf8]{inputenc}
\usepackage[T1]{fontenc}
\usepackage{mathptmx}
\usepackage{etoolbox}
\usepackage{amsmath}

\makeatletter
\def\@email#1#2{%
 \endgroup
 \patchcmd{\titleblock@produce}
  {\frontmatter@RRAPformat}
  {\frontmatter@RRAPformat{\produce@RRAP{*#1\href{mailto:#2}{#2}}}\frontmatter@RRAPformat}
  {}{}
}%
\makeatother
\begin{document}

\preprint{AIP/123-QED}

\title[Rotational Coherence Dominates Early-Time Dynamics and Produces Long-Time Revivals in the $S_2$ State of Azulene]{Rotational Coherence Dominates Early-Time Dynamics and Produces Long-Time Revivals in the $S_2$ State of Azulene}

\author{Jie Zhan}
\affiliation{%
Department of Chemistry, University of Georgia, Athens, GA
}%
\author{Alexander K. Lemmens}%
\author{Musahid Ahmed}%
 \email{MAhmed@lbl.gov}
\affiliation{Chemical Science Division, Lawrence Berkeley National Laboratory, Berkeley, CA
}%
\author{Melanie A. R. Reber}
\email{mreber@uga.edu}
\affiliation{%
Department of Chemistry, University of Georgia, Athens, GA
}%

\date{01/06/2026}

\begin{abstract}
The ultrafast dynamics of azulene have been debated for decades, with reported picosecond decay constants variously attributed to intramolecular vibrational redistribution (IVR), internal conversion, or rotational dephasing. Using polarization and femtosecond time-resolved Resonance Enhanced Multi-photon Ionization Spectroscopy with a nanosecond delay window, we disentangle this long-standing inconsistency and show that the early 2–5 ps decay component arises entirely from rotational dephasing of an excited-state wavepacket. Identical time constants extracted from the decay of the parallel signal and rise of the perpendicular signal across multiple vibronic origins provide an unambiguous rotational anisotropy signature, eliminating the need for IVR-based interpretations. Extending the measurement window to 1.3 ns reveals well-structured J-type and C-type rotational coherence revivals in $S_2$ azulene on top of the well-documented fluorescence decay, demonstrating that both the short- and long-time dynamics contain information about the coherent rotational dynamics. These results show that azulene, and by extension polycyclic aromatic hydrocarbons (PAH), can sustain structured rotational coherence deep into the nanosecond regime, positioning PAHs as model systems for quantum-coherent wavepacket dynamics and providing a framework for probing coherence, decoherence, and rotational control in electronically rich molecular systems. 
\end{abstract}

\maketitle

\section{Introduction}

Azulene is a 10-\(\pi\)-electron aromatic molecule distinguished by its bright \(S_2\) excited state and dark \(S_1\) state, a consequence of its well-known anti-Kasha behavior.\cite{dunlop2023excited} Although its femtosecond dynamics have been extensively investigated in condensed phases,{\cite{wirth1976ultrafast,wurzer2000highly,ciano1997dynamics,solowan2022direct,sawada2025ultrafast} only a limited number of studies{\cite{RAFFAEL200859,blanchet2008time,diau1999direct,demmer1987}  have examined \(S_2\) dynamics in molecular beams with femtosecond to picosecond time resolution. These gas-phase studies report notable differences in both the observed decay behavior and its interpretation.

Azulene’s \(S_2\) dynamics are often described by two principal time constants: one in the picosecond regime and another in the nanosecond regime. In 1987, Demmer et al.{\cite{demmer1987}} reported a 2 ns decay attributed to radiative relaxation of \(S_2\), along with an initial sub-30 ps decay that they assigned to intramolecular vibrational redistribution (IVR). This interpretation was later adopted by Diau et al.{\cite{diau1999direct}} in a femtosecond time-resolved mass-spectrometry experiment, where a 350 fs decay component was observed and described as consistent with the IVR timescale reported by Demmer et al.\cite{demmer1987} A later study by Raffael et al.{\cite{RAFFAEL200859}} argued instead that the picosecond component arises from rotational dephasing, based on time-resolved photoionization measurements comparing parallel and perpendicular pump–probe geometries. They reported dephasing times of 1.7–4.3 ps for seeded versus pure molecular beams, though their data extended only to 3 ps and did not include full kinetic fits. In a companion study, the same group attempted to observe rotational coherence revivals at longer delays but did not detect them.\cite{blanchet2008time}
In this work, we address the inconsistencies in the reported early-time dynamics of \(S_2\) azulene by conducting experiments using pump energies that selectively access \(S_2\) → \(S_0\) vibronic transitions. This approach provides vibrational-state specificity together with femtosecond time resolution extending into the nanosecond regime, allowing us to examine both the origin of the picosecond decay and the presence of rotational coherence revivals in addition to the \(S_2\)- \(S_0\) fluorescence associated with the \(S_2\) population decay. Parallel, perpendicular, and magic angle pump–probe polarization schemes were employed to verify the role of rotational dephasing in the early-time signal. By combining femtosecond excitation with a nanosecond delay stage, we also investigated longer-time dynamics up to 1.3 ns and analyzed rotational coherence in this extended time window for the first time.

\section{Methods}
\subsection{Experimental}

The experiments were conducted using a home-built Time-Resolved Resonance-Enhanced Multiphoton Ionization (TR-REMPI) spectrometer at Lawrence Berkeley National Laboratory (LBNL).\cite{lemmens2025vibrations} Laser pulses from a Coherent Monaco Yb:fiber laser were split into separate pump and probe beams. The probe beam, comprising 33\% of the seed, was frequency-tripled to 345 nm via third-harmonic generation, while the remaining 67\% was directed to a Light Conversion Opera-HP optical parametric amplifier to generate the pump wavelengths. A motorized delay stage on the probe arm enabled scans extending past 1 ns. Pump wavelengths from 348.2 to 329.1 nm (3.56 to 3.77 eV) were used in the experiments. The pump and probe beams were focused and intersected with the molecular beam in the interaction region.

Solid azulene (Sigma-Aldrich, 99\%) was placed in a stainless-steel reservoir heated to 350 K, with 450 Torr of argon flowing over it. The resulting mixture was expanded through a 50 \(\mu\)m nozzle into vacuum to form a supersonic molecular beam. After passing through a 2 mm skimmer, the beam entered the interaction region where it overlapped with the pump and probe. Azulene has an ionization energy of 7.4 eV; to achieve 1+2' ionization, a tightly focused probe beam was used in combination with a weakly focused pump beam. Ions were collected using a microchannel plate (MCP) detector coupled to a time-of-flight mass spectrometer. The signal intensity is obtained by integrating the mass spectrum over a window centered on the target mass. The instrument response function has a time resolution of 350 fs and a 1.2 nm FWHM as determined from the observed minimum linewidths, allowing both frequency-resolved and time-resolved measurements to be obtained.

\subsection{Molecular Structure Calculations}
Calculations of the \(S_0\) and \(S_2\) states of azulene were performed using Gaussian16{\cite{g16}} at the CAM-B3LYP/6-311G(d,p) level of theory. The ground-state geometry was optimized using DFT, and the excited-state analysis was carried out using TD-DFT. All optimizations were constrained to \(C_{2v}\) symmetry. Franck–Condon and Herzberg–Teller (FCHT) analyses were subsequently performed using Gaussian16, and stick spectra were generated from the computed vibronic intensities. The electronic-structure methodology follows the level of theory employed by Palmer et al.{\cite{palmer2022excited}}, while all FCHT calculations and spectral generation were performed independently in this work. The resulting spectra are shown in Figure \ref{fig:fig1}.

\subsection{Time Resolved Data Model}
The time-dependent REMPI signal contains several contributing processes, all of which are included in the kinetic model. These processes are: the instrument response function, a coherent artifact, a fast decay component, and a slow decay component. For delays up to 20 ps, the kinetics for both parallel and perpendicular pump–probe polarizations are adequately described by a single-exponential process. The model consists of a Gaussian term with width, \(\omega_\delta\), that accounts for the coherent artifact (CA), followed by a fast component representing the dynamics within the first 20 ps with a time constant $\tau$. This fast component appears as a single-exponential decay for the parallel configuration and as a corresponding rise for the perpendicular configuration, which represents a phenomenological fit as the rotational dephasing is not strictly exponential. A slow component associated with fluorescence with a lifetime exceeding 1 ns, is also included; with $S$ and $d$ as the fit parameters. The parameter \(t_0\) denotes the temporal overlap of the pump and probe pulses and defines time zero and each component has an intensity scaling factor, $P_n$.
  \begin{align}
&\text{CA} =
\frac{P_{1}}{\sqrt{2\pi}\,w_{\delta}}
\exp\!\left[-\frac{(t - t_{0})^{2}}{2 w_{\delta}^{2}}\right] \\
&\text{Fast Component (Parallel)} = P_2 e^{-(t - t_0)/\tau}, \notag\\
&\qquad\text{for } t \ge t_0 \\
&\text{Fast Component (Perpendicular)} = P_2 (1 - e^{-(t - t_0)/\tau}), \notag\\
&\qquad\text{for } t \ge t_0 \\
&\text{Slow Component} = S\times(t - t_0) + d, \notag\\
&\qquad\text{for } t \ge t_0
\end{align}

The total kinetics model is therefore: 
    \begin{align}
    &\text{Total Signal} =
\left( \text{CA} + \text{Fast Component} + \text{Slow Component} \right),
\notag\\
&\qquad \text{for } t \ge t_0
\end{align}

In ultrafast pump–probe measurements, the minimum achievable time resolution is determined by the instrument response function (IRF), which quantifies the combined temporal response of the excitation and detection system. The measured IRF is well described by a Gaussian function with a width, $\omega_{irf}$: 
    \begin{align}
    \mathrm{IRF}(t, \omega_{irf}) =
    \frac{1}{\sqrt{2\pi}\,\omega_{\mathrm{irf}}}
    \exp\!\left[-\frac{(t - t_0)^2}{2\omega_{\mathrm{irf}}^2}\right].
    \label{eq:irf_t0}
    \end{align}

The raw data are fit to the convolution of the kinetic model and the IRF (Eq. 7). 

    \begin{align}
    \begin{split} 
    I(t) &= \Bigg\{CA + signal \cdot H(t - t_0)\Bigg\} \otimes \mathrm{IRF}(t,\omega_{irf}), \\[6pt]
    &\quad
    H(t - t_0) =
    \begin{cases}
    1, & t \ge t_0,\\
    0, & t < t_0.
    \end{cases}
    \end{split}
    \label{eq:parallel}
    \end{align}

The Heaviside step function \( H(t-t_0) \) ensures that the kinetic terms contribute only for \(t \geq t_0\). All time-dependent data were fit to this model using a non-linear least-squares routine.

\section{Results and Discussion}

    \begin{figure}[ht]
    \centering
    \fbox{\includegraphics[width=0.5\textwidth]{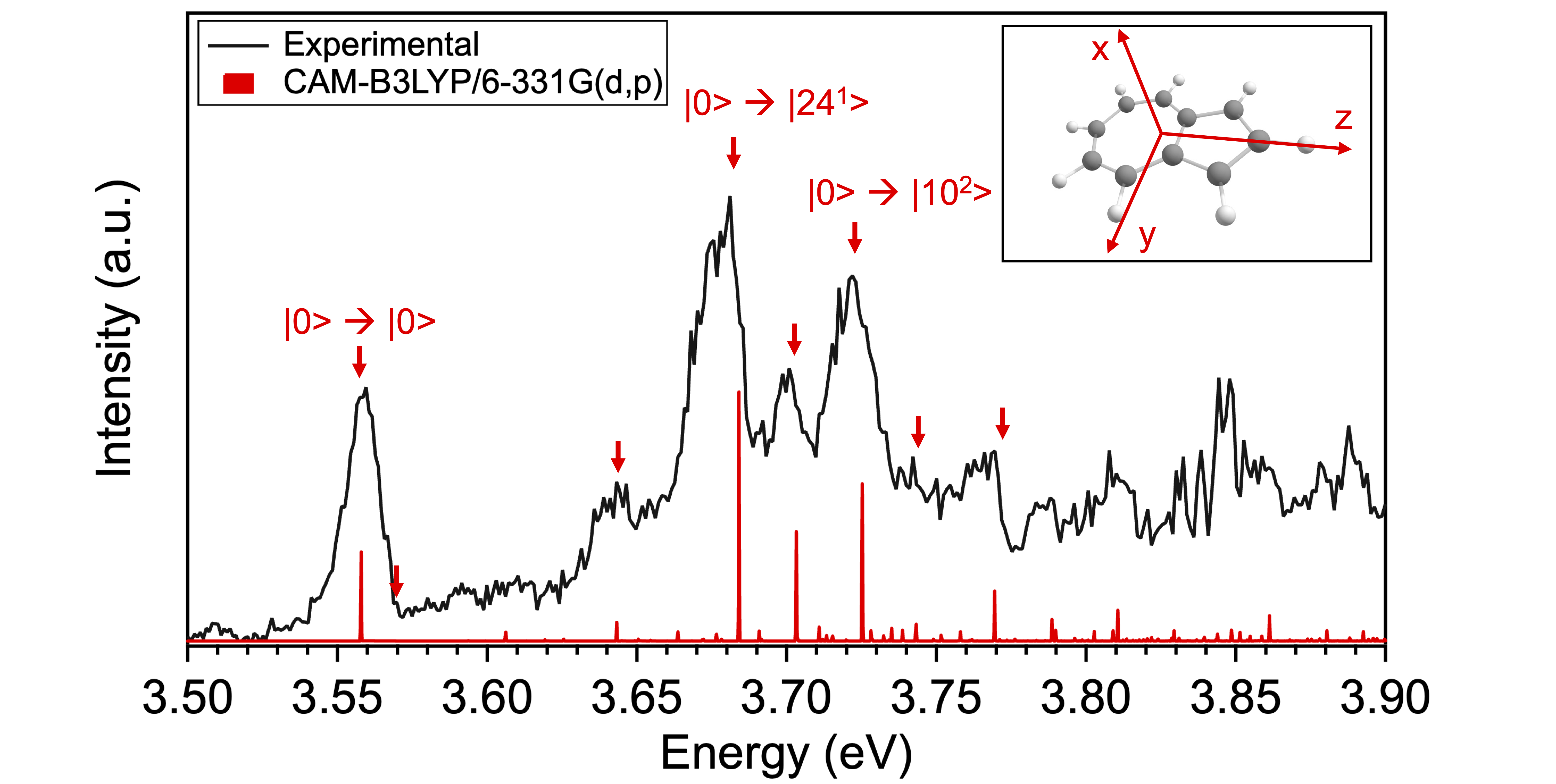}}
    \caption{Measured REMPI Spectrum and calculated Franck-Condon Herzberg-Teller spectrum at 0 K. The FWHM of the calculated spectrum is 1 \(cm^{-1}\) and shifted to a lower energy by 0.091 eV. This work specifically studied the photon energies indicated by the red arrows. The inset shows the structure and principle axes of azulene.}
    \label{fig:fig1}
    \end{figure}

Figure \ref{fig:fig1} shows the 1+2' REMPI spectrum together with the simulated Franck–Condon and Herzberg–Teller (FC–HT) stick spectrum of azulene. The probe photon energy was fixed at 3.59 eV while pump photon energies were scanned from 3.5 to 3.9 eV. The calculated spectrum was generated at 0 K and shifted by \(-0.091~\text{eV}\) to align with the experimental results, showing good agreement between experiment and theory. Vibronic assignments follow the notation \(|S_0~vibrational~state> \rightarrow |S_2~vibrational~state>\), with the vibrational quanta indicated as superscripts.

The origin of the experimental (\(|0> \rightarrow |0>\) ) transition occurs at 3.561 eV (28719~\(\mathrm{cm}^{-1}\), 348.2 nm), in good agreement with the previously reported value of 28757~\(\mathrm{cm}^{-1}\) (3.565 eV, 347.7 nm).{\cite{demmer1987}} Based on the REMPI spectrum, eight pump photon energies corresponding to distinct vibronic transitions were selected for time-resolved measurements under the 1+2' scheme. Photon energies above 3.77 eV were excluded because the sum of a pump and probe photon exceeds the ionization energy of azulene, placing those conditions outside the 1+2' regime. Ultrafast REMPI experiments were therefore conducted at 3.56, 3.57, 3.64, 3.68, 3.70, 3.72, 3.74, and 3.77 eV, as indicated by the red arrows in Figure 1. Pump  Seven of these energies are resonant with $|S_0> \rightarrow |S_2>$ vibronic transitions, while one (3.57 eV) is off-resonance. The selected energies sample both isolated vibronic bands and regions with overlapping transitions.

    \begin{figure}[ht]
    \centering
    \fbox{\includegraphics[width=0.45\textwidth]{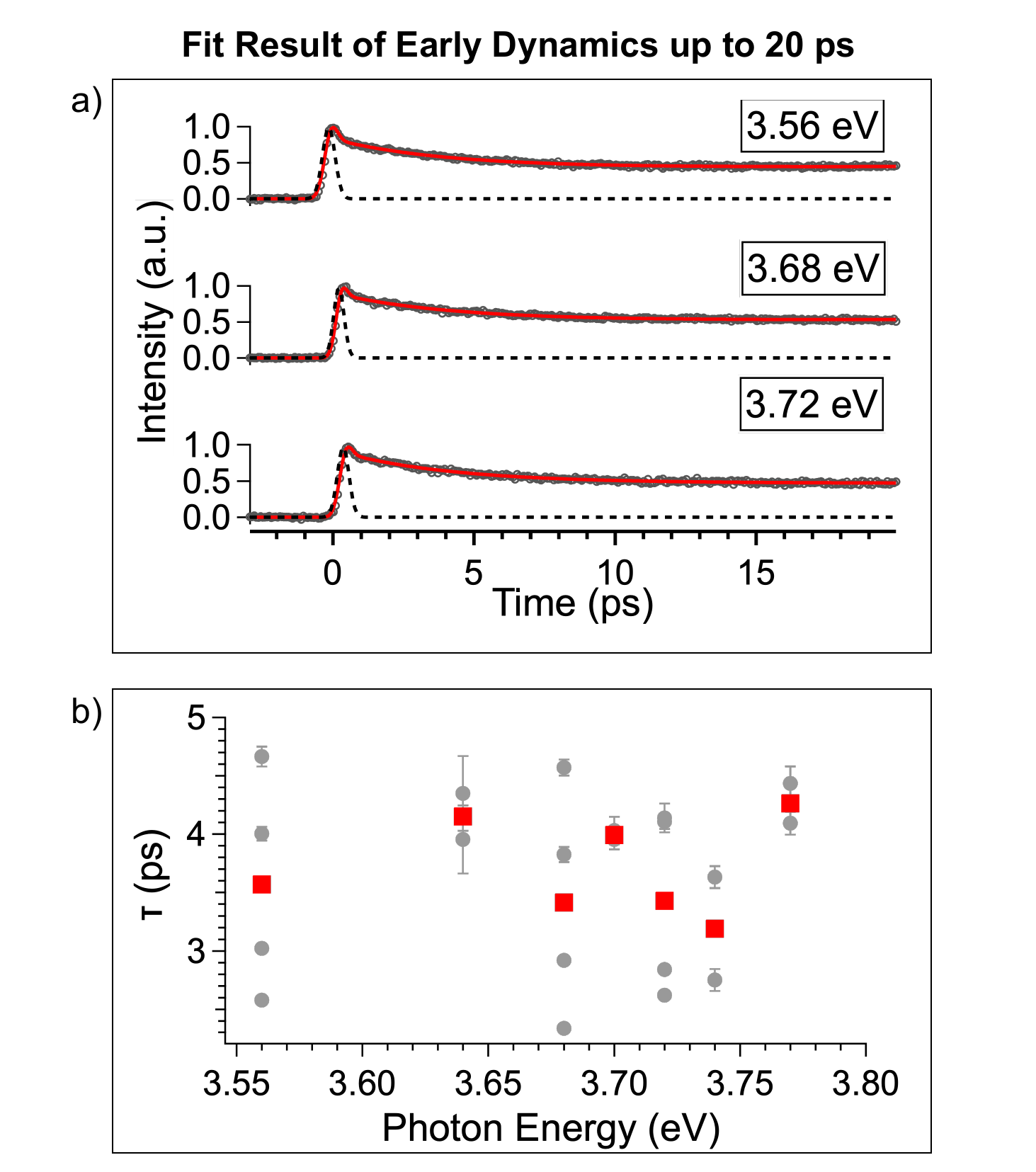}}

    \caption{a) Normalized decay kinetics (grey markers) measured for early time dynamics with pump energy of 3.56 eV, 3.68 eV and 3.72 eV. Red curve shows exponential fit result and the black dashed traces are the fitted IRF. b) Time constants from the fit results for all seven pump energies (grey circles with error bars), The average time constants of the early time dynamics (red squares).}
    \label{fig:fig2}
    \end{figure}

Figure \ref{fig:fig2}a shows single-exponential fits to the early-time decay at 3.56, 3.68, and 3.72 eV under parallel pump–probe polarization. The early time decay of off-resonance pump energy (3.57 eV) is showed in Figure S1. Figure \ref{fig:fig2}b summarizes the extracted fast-decay time constants for all pump energies; complete fit results are provided in the Supporting Information (Table S1). Gray circles represent individual fitted time constants with their corresponding uncertainties in the fit, while the red squares denote the average value at each wavelength. This early-time constant is attributed to the initial dephasing of rotational coherence.{\cite{FemtoChem_Chapter5}}

When molecules are excited by a polarized pulse, those with transition dipole moments aligned along the polarization axis are preferentially excited. As the molecules freely rotate, this initial alignment rapidly dephases, giving rise to the picosecond-scale decay. The intensity of the rotational coherence signal I depends on time, the excitation and detection polarizations, the orientation of the relevant transition dipole moments, and the rotational temperature of the sample.{\cite{zewail1987_theory}} No systematic trend in the extracted time constants was observed for photon energies above the vibronic origin. Additional evidence supporting the rotational-dephasing assignment comes from the perpendicular pump–probe configuration, in which the signal exhibits a rise rather than a decay (Figure \ref{fig:fig3}a). The fit results for the parallel and perpendicular pump-probe configurations at the same molecular beam parameters are provided in Table S2. Measurements were also performed using the magic-angle pump–probe configuration (Figure S2) to confirm the absence of decay mechanisms other than rotational coherence. This behavior is characteristic of rotational dynamics and has been observed previously by Raffael et al.;{\cite{RAFFAEL200859}} however, they reported a sub-1-ps dephasing time, likely due to the higher rotational temperature of their molecular beam.{\cite{FemtoChem_Chapter5}} 

    \begin{figure}[htbp!]
    \centering
    \fbox{\includegraphics[width=0.5\textwidth]{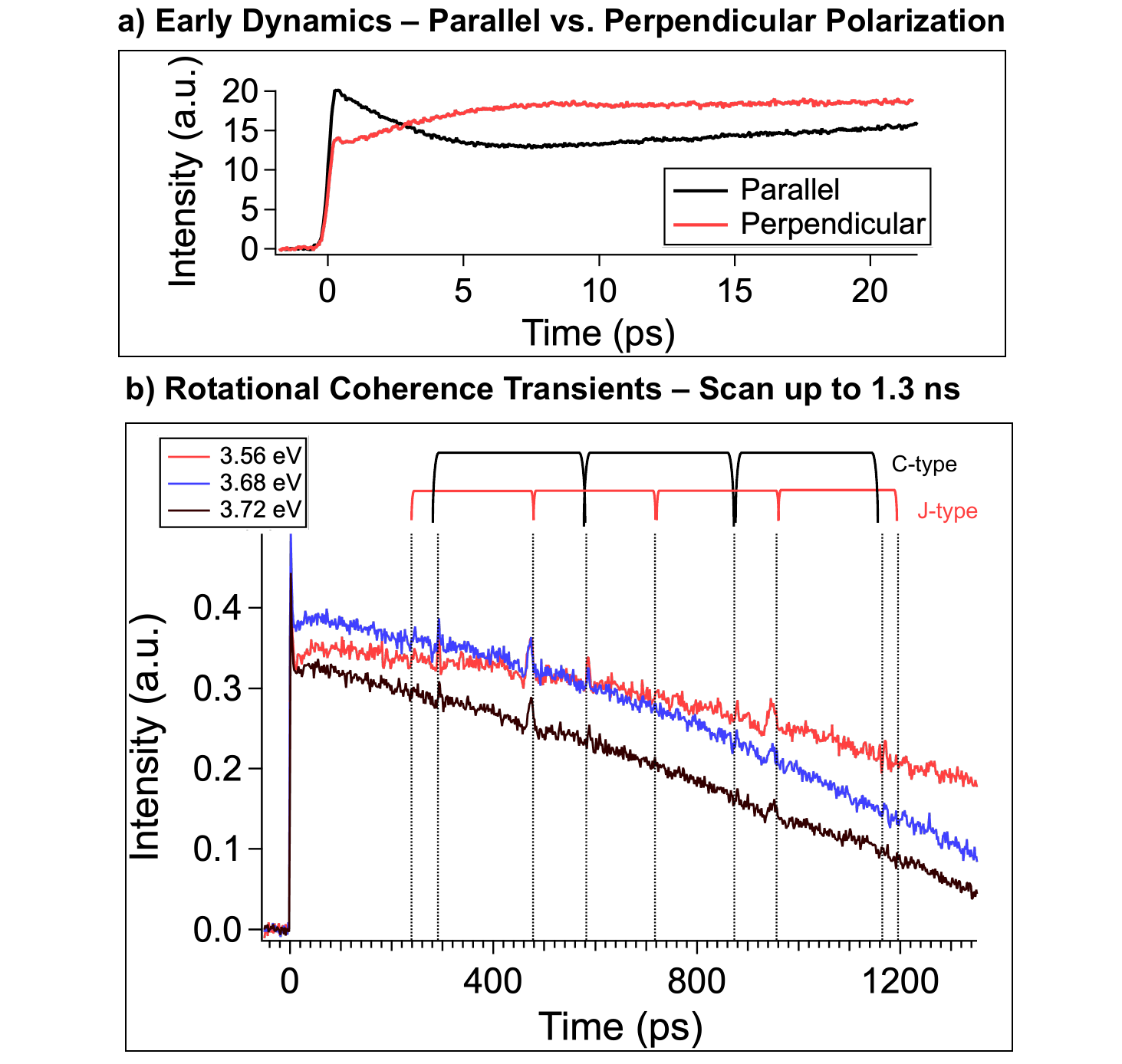}}

    \caption{a) Decay kinetics pumped at 3.56 eV with different pump polarization. b) Normalized Time-Resolved REMPI spectra pumped at 3.56, 3.68 and 3.72 eV scanned up to 1.3 ns with J type and C type rotational coherence transients labeled. J type transients labeled in red indicating a positive signal and blue indicating a negative signal. The results are calculated within the near-prolate top approximation.}
    \label{fig:fig3}
    \end{figure}

We also calculated the rotational anisotropy (Figure S3) for the initial several picoseconds at pump photon energies of 3.56, 3.78, and 3.72 eV. In all three cases, the anisotropy has a value of approximately 0.2 at time zero, consistent with the results reported by Raffael et al.{\cite{RAFFAEL200859}} and significantly lower than the theoretical maximum of 0.4.{\cite{zewail1987_theory}} This reduction in anisotropy is physically meaningful. Thermal averaging over multiple rovibronic transitions, each with different transition dipole orientations, would reduce the initial anisotropy. An additional contribution may arise from a mismatch between the vibronic transition dipole addressed by the pump and the photoionization dipole that governs the probe, which would also prevent the anisotropy from reaching its theoretical limit of 0.4.{\cite{scherer1987sub}}

To further confirm the identity of the initial decay constant and to test whether any part of it could be attributed to IVR, we extended the ultrafast REMPI measurements to delays of 1.3 ns, as shown in Figure \ref{fig:fig3}b. The three scans correspond to pump photon energies of 3.56, 3.68, and 3.72 eV, with a step size of 3 ps. The long-time decay is characterized by a time constant of 1482 ± 73 ps, which we assign to the \(S_2\) fluorescence lifetime, in agreement with the value reported by Demmer et al.{\cite{demmer1987}} At extended nanosecond delays, we also observe a recurring signal associated with rotational coherence in \(S_2\) azulene, which is reported here for the first time.

Azulene is a near-prolate asymmetric top, characterized by Ray's asymmetry parameter \(\kappa = -0.61\). Our analysis therefore follows previous treatments of near-symmetric top species.\cite{FemtoChem_Chapter5} The \(S_2\) state of azulene exhibits rotational constants that satisfy \((B + C) \approx A \gg (B - C)\) as expected for a near-prolate top. For the dominant vibronic transitions investigated here, the transition dipole moment lies along the \textit{a} axis, also denoted as the \textit{z} axis and taken as the principal axis of the molecule. On this basis, we assign the labeled features in the time-resolved data as J-type and C-type transients.{\cite{joireman1992characterization}}

J-type transients can arise for nearly any orientation of the transition dipole moment relative to the principal axes. They correspond to coherences between states with different total angular momentum but the same projection quantum number ($ |\Delta J| = 1,2~~|\Delta K|=0$).{\cite{felker1992rotational}} These transients persist even as the rotor becomes asymmetric. For asymmetric molecules such as azulene, J-type revivals occur at approximately \(t = 1/2(B+C)\) and exhibit alternating signal polarity.{\cite{joireman1992characterization}} 

C-type transients, by contrast, are “asymmetry transients”. They arise from $\Delta J = 2,~\Delta K=0$ rovibronic coherences{\cite{felker1992rotational}} and appear sharper in the time domain than the main J-type peaks.{\cite{joireman1992characterization}} C-type transients are characteristic of prolate-like species when the transition moment lies along the \textit{a} axis, which applies to all three vibronic transitions excited in this experiment. The observed transients therefore show a strong correlation with the transition dipole moment of the molecule. For the 3.56 eV (\(|0> \rightarrow |0>\) and 3.68 eV (\(|0> \rightarrow |24^1>\)) transitions, the transition dipole moment derivative is oriented along the \textit{a} axis, whereas the 3.72 eV transition \(|0> \rightarrow |10^2>\) has its transition dipole moment derivative along the \textit{c} axis (\textit{x} axis). As a result, the effective transition dipole moments for these three bands are distributed between the \textit{a} and \textit{c} axes. C-type transients are revived at approximately \(t = 1/4C\).{\cite{joireman1992characterization}} 

    \begin{table}[h]
\centering
\caption{\bf Revival times for J-type and C-type signals (ps)}
\label{tab:revivals}
\begin{tabular}{c|l}
\hline
\textbf{Type} & \textbf{Revival times (ps)} \\
\hline
J-type & 239.2,\; 478.3,\; 717.5,\; 956.7,\; 1195.8 \\
C-type & 291.2,\; 582.4,\; 873.6,\; 1164.8 \\
\hline
\end{tabular}
\end{table}

The J-type and C-type transients of excited-state azulene were calculated using the \(S_2\) rotational constants from Semba et al.,{\cite{semba2009rotationally}} following the approach outlined in Ref. 14. The calculated revival times are listed in Table \ref{tab:revivals}. For the three pump energies shown in Figure \ref{fig:fig3}b, the recurring experimental features appear at approximately the same time delays. This indicates that the rotational constants of the final vibrational states accessed at these energies are very similar to those of the \(S_2\) origin, and that their geometries are therefore comparable. At 3.72 eV, the rotational coherence revivals display a weaker C-type contribution, consistent with the transition dipole moment not being purely aligned along the a axis. No A-type transients are observed, implying that the derivative of the transition dipole moment along the remaining principal axis makes only a minor contribution to the total transition dipole. The J-type revival times at the three pump energies are observed to shift to slightly earlier delays, as expected for asymmetric tops.{\cite{FemtoChem_Chapter5}} For asymmetric-top molecules, J-type transients at \(t = (2m +1)/2(B+C),~ (m=0,1,2 ...) \) lose intensity more rapidly,\cite{zewail1987_theory,joireman1992characterization,baskin1987purely_exp} in agreement with our observations. With these assignments in hand, we use the widths of the J-type and C-type revivals to model the initial decay.\cite{Zhan_inpreparation} The early-time decay can be reproduced by summing the dephasing of both sets of transients (Figure S4), ruling out IVR as the decay mechanism.

\section{Conclusion}

The present study provides a unified interpretation of \(S_2\)-state dynamics in azulene by combining polarization-resolved femtosecond TR-REMPI measurements with extended nanosecond detection. The widely reported 2–5 ps decay observed in  earlier studies is shown here to originate from rotational dephasing rather than intramolecular vibrational redistribution. This assignment is supported by the identical time constants extracted from the decay of the parallel signal and the rise of the perpendicular signal across several vibronic transitions, which together constitute a clear rotational anisotropy signature. This conclusion is further confirmed by the presence of rotational revivals at long times. This result resolves long-standing inconsistencies in the azulene literature, where variations in beam temperature and excitation conditions produced apparently conflicting ''IVR timescales.''

Extending the measurement window to 1.3 ns reveals structured rotational coherence in \(S_2\) azulene, manifested as J-type and C-type transient features whose timings are consistent with those expected for a near-prolate asymmetric top. The recurrence structure connects continuously to the early-time anisotropy decay, indicating that the short- and long-time behaviors arise from a single underlying rotational coherence. Differences in the strength of C-type contributions among the probed vibronic bands are explained by the orientation of their transition dipole moment derivatives, highlighting the sensitivity of rotational coherence to vibronic character.

These findings establish that excited-state azulene supports rotational coherence over hundreds of picoseconds and into the nanosecond regime, behavior not previously documented for this prototypical aromatic chromophore. This refined picture of \(S_2\) dynamics clarifies the mechanistic origin of both the early picosecond decay and the late-time recurrence structure, and it provides a framework for reinterpreting prior ultrafast measurements on azulene and related PAHs.

Looking ahead, the observation of long-lived rotational coherence in an electronically excited PAH suggests several promising directions. Measurements across additional vibronic transitions could map structural changes of vibronic states along with associated changes in lifetimes or rotational dephasing dynamics. Complementary modeling of asymmetric-top rotational dynamics, or the use of temperature-controlled molecular beams, could quantify the role of rotational temperature and asymmetry splitting in shaping the observed transients. Finally, extending similar measurements to larger PAHs would test how molecular size, rigidity, and $\pi$-electron delocalization influence the persistence and structure of rotational coherence.

\begin{acknowledgments}
SL \& MA are supported by the Gas Phase Chemical Physics Program, in the Chemical Sciences Geosciences and Biosciences Division of the Office of Basic Energy Sciences of the U.S. Department of Energy under Contract No. DE-AC02-05CH11231. JZ \& MR acknowledge support from the National Science Foundation CAREER Program under Award Number 2340180.
\end{acknowledgments}

\section*{Data Availability Statement}

The data that support the findings of this study are available from the corresponding author upon reasonable request.

\section*{Credit} 

The following article has been submitted for peer-review publication. After it is published, it will be found at the journal website.

\section*{References}

\bibliography{aipsamp}

\pagebreak
\newpage
\widetext
\begin{center}
\textbf{\large Supplemental Materials}
\end{center}
\setcounter{equation}{0}
\setcounter{figure}{0}
\setcounter{table}{0}
\setcounter{page}{1}
\makeatletter
\renewcommand{\theequation}{S\arabic{equation}}
\renewcommand{\thefigure}{S\arabic{figure}}
\renewcommand{\bibnumfmt}[1]{[S#1]}
\renewcommand{\citenumfont}[1]{S#1}

\newpage 

\section{Additional Figures}


\begin{figure}[h]
    \centering
    \includegraphics[width=\textwidth]{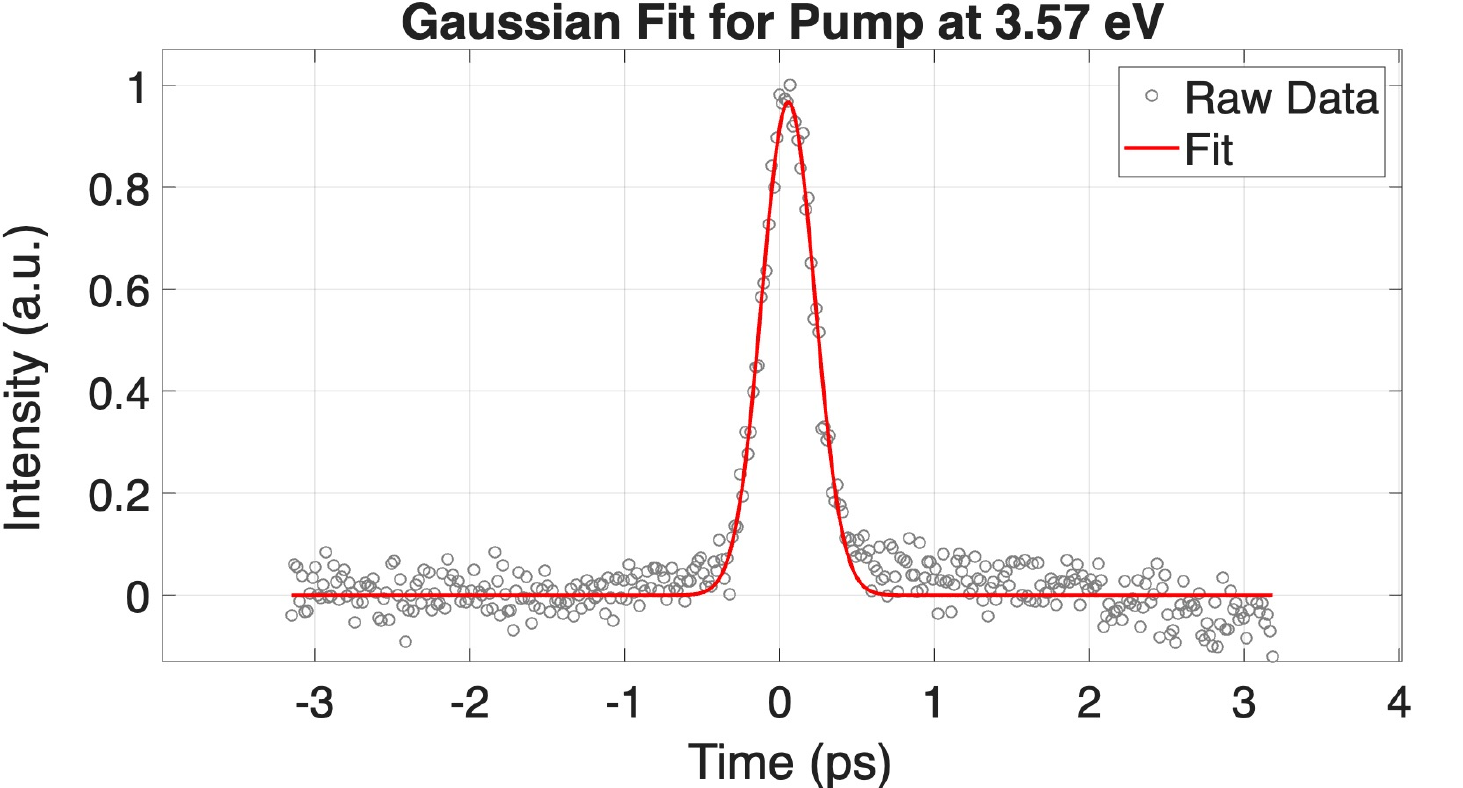}
    \caption{Gaussian fit for early time dynamics for $S_2$ azulene when pumped at 3.57 eV. The fitted width is 
    $0.1712 \pm 0.0049~\mathrm{ps}$, corresponding to an FWHM of 0.4032~ps. This FWHM lies within the expected IRF range, indicating the absence of measurable absorption at off-resonance pump energies and demonstrating that the experiment achieves vibronic-state–selective resolution.}
\end{figure}

\begin{figure}[h]
    \centering
    \includegraphics[width=\textwidth]{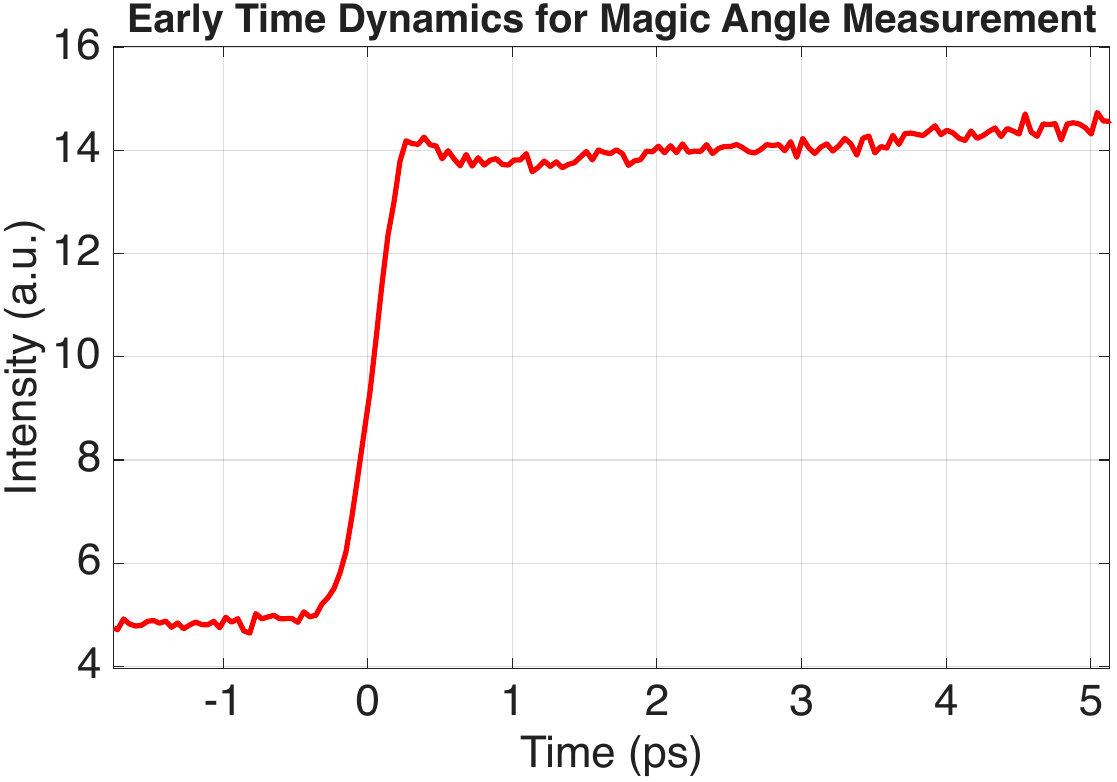}
    \caption{Raw Data of early time dynamics at magic angle. Pumped at 3.56 eV.}
\end{figure}

\begin{figure}[h]

    \centering
    \includegraphics[width=\textwidth]{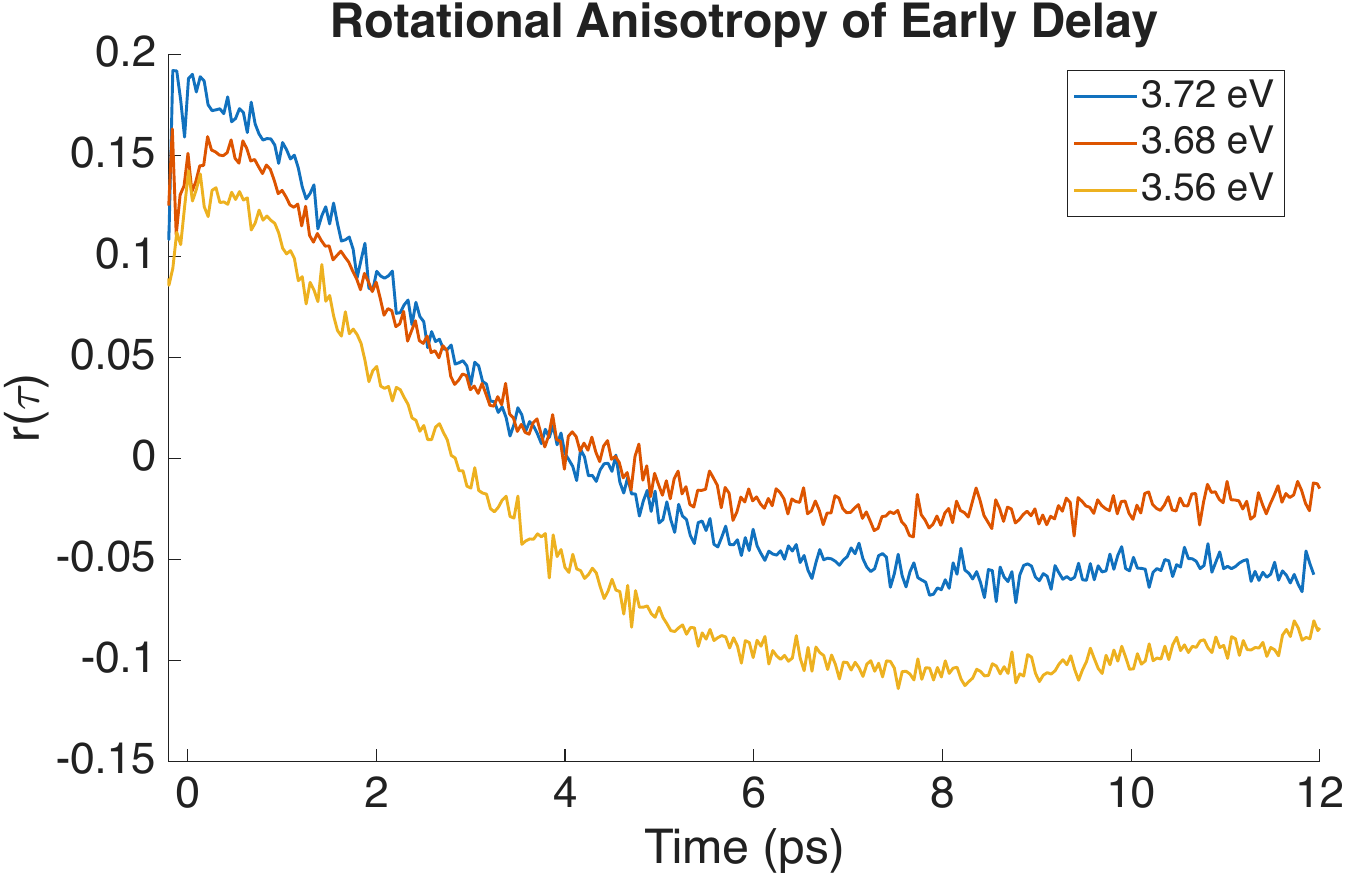}
    \caption{Rotational anisotropy for pump photon energy of 3.72 eV, 3.68 eV and 3.56 eV. Rotational anisotropy is defined by the following equation: $ r(\tau) = \frac{I_{\parallel}(\tau) - I_{\perp}(\tau)} {I_{\parallel}(\tau) + 2 I_{\perp}(\tau)}$, where $r(\tau)$ is the anisotropy at a certain time delay, while $(I_{\parallel}$ and $I_{\perp}$ represent the magnitude of the signal with parallel or perpendicular pump polarization, respectively.  The maximum value of initial anisotropy $r(0)$ could range from 0.4 to -0.2, depending on the angle between the excitation and detection transition's dipole moments, which can vary from being parallel to perpendicular to each other.}
    
\end{figure}

\begin{figure}[h]
    \centering
    \includegraphics[width=\textwidth]{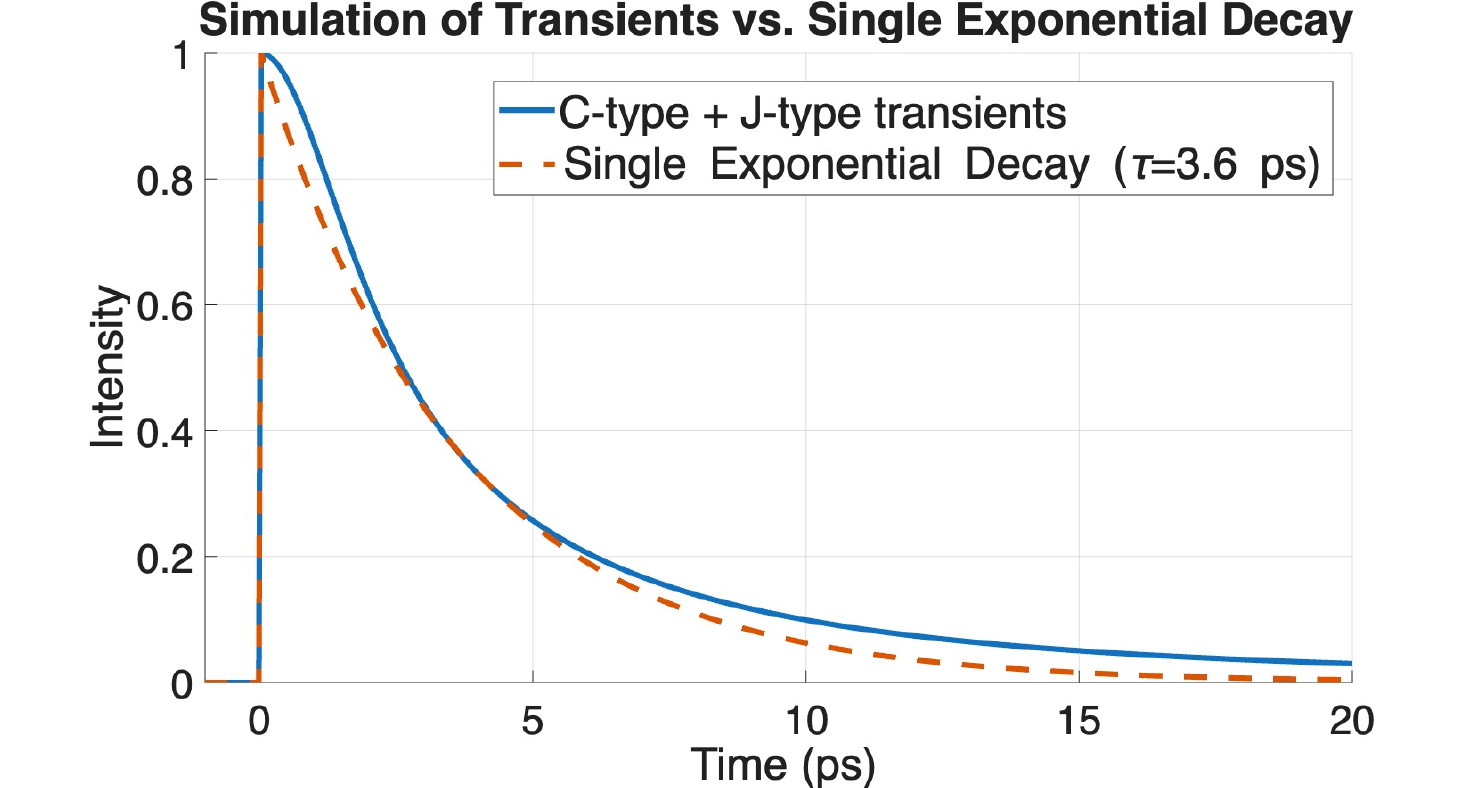}
    \caption{Simulation of the initial C-type plus J-type rotational coherence dephasing model versus single exponential decay model at early time delay.}
\end{figure}

\clearpage
\section{Supplementary Tables}

\begin{table}[!htbp]
\caption{Fitting parameters for scans up to $20~\mathrm{ps}$ using the time-resolved data model Eqs. (1)–(7). The parameter $\omega_{\delta}$ represents the width of the coherent artifact (CA). $P_{2}$ denotes the amplitude of the initial rotational coherence dephasing term, and $\tau$ is the corresponding dephasing lifetime. $\omega_{\mathrm{irf}}$ represents the fitted instrument response function (IRF) width obtained from the time-resolved model. For photon energies marked with $*$, the reported errors represent half of the difference between the largest upper and the smallest lower bounds of the 95$\%$ confidence interval; for all other photon energies, the errors correspond to standard errors.}
\label{tab:fit_params1}
\resizebox{\textwidth}{!}{%
\begin{tabular}{c c c c c c c c c}
\hline
\hline
Photon Energy (eV) & Avg. $\omega_{\delta}$ (ps) & error & $P_2$ & error & Avg. $\tau$ (ps)& error & Avg. $\omega_{IRF}$ (ps) & error \\
\hline
3.56 & 0.0243 & 0.0090 & 0.4733 & 0.025 & 3.5669 & 0.471 & 0.1727 & 0.009 \\
3.64$*$ & 0.0119 & 0.0017 & 0.1755 & 0.025 & 4.1508 & 0.503 & 0.1954 & 0.005 \\
3.68 & 0.0098 & 0.0029 & 0.3947 & 0.010 & 3.4126 & 0.492 & 0.1749 & 0.006 \\
3.70$*$ & 0.0109 & 0.0011 & 0.3487 & 0.035 & 3.9916 & 0.139 & 0.1776 & 0.009 \\
3.72 & 0.0711 & 0.0406 & 0.4385 & 0.022 & 3.4266 & 0.404 & 0.1595 & 0.010 \\
3.74 & 0.0388 & 0.0334 & 0.3738 & 0.014 & 3.1911 & 0.534 & 0.1724 & 0.005 \\
3.77$*$ & 0.0194 & 0.0082 & 0.3713 & 0.026 & 4.2631 & 0.293 & 0.1994 & 0.005 \\
\hline
\end{tabular}}
\end{table}

\begin{table}[!htbp]
\caption{ Fit parameters for measurements acquired with parallel or perpendicular pump--probe 
polarization schemes up to $20~\mathrm{ps}$, analyzed using the time-resolved data model 
Eqs. (1)–(7). Here, $\parallel$ denotes parallel pump--probe polarization, and $\perp$ 
denotes perpendicular pump--probe polarization. The parameter $\omega_{\delta}$ represents the width of the coherent artifact (CA). $P_{2}$ denotes the amplitude of the initial rotational coherence dephasing term, and $\tau$ is the corresponding dephasing lifetime. $\omega_{\mathrm{irf}}$ represents the fitted instrument response function (IRF) width obtained from the time-resolved model. The reported errors represent half of the difference between the
upper and lower bounds of the 95$\%$ confidence interval.}
\label{tab:fit_params2}
\resizebox{\textwidth}{!}{%
\begin{tabular}{c c c c c c c c c c}
\hline
\hline
Polarization & Photon Energy (eV) & $\omega_{\delta}$ (ps) & error & $P_2$ & error & $\tau$ (ps) & error & $\omega_{\mathrm{irf}}$ (ps) & error\\
\hline
$\parallel$        & 3.56 & 0.0078 & 0.000 & 0.5021 & 0.002 & 2.5791 & 0.032 & 0.1630 & 0.002 \\
$\perp$   & 3.56 & 0.0084 & 0.000 & 0.3352 & 0.002 & 2.8660 & 0.060 & 0.2151 & 0.003 \\
\hline
$\parallel$        & 3.68 & 0.0058 & 0.000 & 0.3895 & 0.002 & 2.3369 & 0.034 & 0.1728 & 0.002 \\
$\perp$   & 3.68 & 0.0140 & 0.000 & 0.2474 & 0.002 & 3.4493 & 0.086 & 0.1830 & 0.003 \\
\hline
$\parallel$        & 3.72 & 0.0065 & 0.000 & 0.4534 & 0.002 & 2.6208 & 0.028 & 0.1427 & 0.001 \\
$\perp$   & 3.72 & 0.0171 & 0.000 & 0.2359 & 0.005 & 3.1201 & 0.171 & 0.1928 & 0.003 \\
\hline
\end{tabular}}
\end{table}

\end{document}